\documentclass[prl,letterpaper,twocolumn,showpacs]{revtex4}
\usepackage{times,xspace}
\usepackage{amsbsy,amssymb,amsmath,bm}
\usepackage{graphicx,color,epsfig}
\usepackage{fancyhdr}

\begin{document}

\title{Gutzwiller density functional theory for correlated electron systems}
\author{K. M. Ho, J. Schmalian, and C. Z. Wang}
\affiliation{Ames Laboratory-U.S. DOE and Department of Physics and Astronomy, Iowa State
University Ames, IA 50011, USA}
\date{\today }

\begin{abstract}
We develop a new density functional theory (DFT) and formalism for
correlated electron systems by taking as reference an interacting electron
system that has a ground state wavefunction which obeys exactly the
Gutzwiller approximation for all one particle operators. The solution of the
many electron problem is mapped onto the self-consistent solution of a set
of single particle Schr\"{o}dinger equations analogous to standard DFT-LDA
calculations.
\end{abstract}

\maketitle

Over the last several decades, first principles total energy calculations
using density functional theory based on the local density approximation
(LDA) or generalized gradient approximation (GGA) have been well developed
into a theoretical tool with strong predictive capability for a large number
of materials\cite{Hohenberg64,Kohn65,Louie96,Vasp}. However, there are
important classes of materials involving strongly correlated electrons
ranging from high Tc superconducting compounds and various other transition
metal oxide materials to f-electron elements bearing materials where the
current LDA/GGA approaches fail in fundamental ways. There have been
intensive studies on new approaches to remedy the situation such as LDA+U%
\cite{Anisimov91}, LDA-dynamical mean field theory (DMFT)\cite%
{Kotliar04,Kotliar06,McMahan05,Savrasov01} , self-interaction
correction-local spin density (SIC-LSD)\cite{Petit02}, and hybrid functionals%
\cite{Kudin02,Prodan05}. Although these approaches partially address the
issues related to the strongly correlated electron systems, a comprehensive
and generally accepted predictive theory with the quality of LDA for normal
metal, alloys, and compounds is still lacking for materials containing
strongly correlated electrons. To address this problem, we propose in this
paper a new density functional theory that goes beyond LDA through a
self-consistent solution of the many-body ground state using the Gutzwiller
approximation\cite{Gutzwiller63} for interacting electron systems within a
first-principles framework.

The Hohenberg-Kohn theorem states that the ground state energy of an
electron system is a functional of the electron density\cite{Hohenberg64}.
Kohn and Sham\cite{Kohn65} took this a step further by expressing the energy
of a real system in terms of the energy of a fictitious system of
non-interacting electrons that has the same density and same kinetic energy
as the real system. This led to a system of non-interacting electrons moving
in an effective potential that can be solved through the iterative solution
of a set of one-electron Schr\"{o}dinger equations within the LDA for the
exchange-correlation energy. The current first-principles density functional
calculations are based on this set of effective one-electron Kohn-Sham
equations. It should be noted that the Hohenberg-Kohn density functional
theorem is true for any electron system including strongly correlated
electron systems. The failure of the Kohn-Sham approach for the strongly
correlated electron systems, in our view, is largely due to the choice of
non-interacting electron as the reference system. By carefully choosing a
reference system which includes the most essential strong correlations, yet
still can be cast into a set of one-electron Schr\"{o}dinger equations
through variational principles, a new density-functional formalism for
treating strongly correlated electron systems can be derived and implemented
following the spirit of Kohn and Sham\cite{Kohn65}.

In the new approach we propose here, instead of defining the kinetic energy
functional to be the kinetic energy of a non-interacting electron gas with
the same density, we will define the kinetic energy functional to have a
simple analytical form corresponding to the frequently used Gutzwiller
approximation for a system of electrons with on-site-only correlations\cite%
{Gutzwiller63,Vollhardt84,Kotliar86,Dorin,Julien05,Buenemann}. The
expectation of any one particle operator (e.g. the electron kinetic energy
and the electron density) under the Gutzwiller approximation can be
expressed in terms of a non-interacting one-particle density matrix with
renormalized weight due to adjustments from strong correlation effects\cite%
{Gutzwiller63,Vollhardt84,Kotliar86,Dorin,Julien05,Buenemann}. We will show
that. within the Gutzwiller approximation, the interacting many-electron
problem can be mapped onto a non-interacting system with an effective
potential. The exact Coulomb interactions can be included for a
pre-determined set of localized configurations while the local density
approximation is used for all the remaining exchange-correlation
contributions.

The choice of including electron correlations using the Gutzwiller
approximation is motivated by previous work where it has been shown to
interpolate well between the two regimes of strong electron correlation
(large U limit) and small electron correlation (small U limit). Another
importance of the Gutzwiller approach is the correct description of highly
correlated states near the Fermi level. LDA can be view as an extension of
Hartree-Fock theory into density functional theory, we view our present
scheme as an extension of quantum chemical couple-cluster calculations into
density functional theory. In our scheme, the variational parameters are the
single particle electron wavefunctions for the localized and delocalized
electrons and the occupancy of the various localized configurations at each
atom in the unit cell. Like LDA, the formulation is from first principles
with all Coulomb and exchange interactions determined self-consistently.
There are no adjustable parameters.

According to the Hohenberg-Kohn density functional theorem, the ground state
energy of a multi-electron system is a functional of the electron density $%
\rho $%
\begin{equation}
\left\langle \Psi \left\vert \widehat{H}\right\vert \Psi \right\rangle =E%
\left[ \rho \right] .
\end{equation}%
Instead of taking as reference a non-interacting electron gas with the same
density as the exact many-electron system, the density functional in our
present theory is determined by taking as reference an \emph{interacting}
electron system:%
\begin{eqnarray}
\left\langle \Psi \left\vert \widehat{H}\right\vert \Psi \right\rangle
&=&\left\langle \Psi _{G}\left\vert \widehat{T}+\widehat{V}_{\mathrm{ion}%
}\right\vert \Psi _{G}\right\rangle +E_{xc}\left[ \rho \right] .  \notag \\
&&+\frac{1}{2}\int \rho \left( \mathbf{r}\right) v\left( \mathbf{r,r}%
^{\prime }\right) \rho \left( \mathbf{r}^{\prime }\right)
d^{3}rd^{3}r^{\prime }  \label{funct}
\end{eqnarray}%
The reference system is chosen to have the same electron density $\rho
\left( \mathbf{r}\right) $ as the ground state of the exact multi-electron
system and to have a ground state wavefunction $\left\vert \Psi
_{G}\right\rangle $ which obeys exactly the Gutzwiller approximation for all
one particle operators%
\begin{equation}
\widehat{O}=\sum_{m=1}^{N}\widehat{O}_{m}
\end{equation}%
In the Gutzwiller approximation, for each one-particle operator acting on $%
\left\vert \Psi _{G}\right\rangle $, we can define a corresponding
renormalized operator $\widehat{O}_{G}$, acting on the underlying
Hartree-like wavefunction $\left\vert \Psi _{0}\right\rangle $used in
generating $\left\vert \Psi _{G}\right\rangle $ in the Gutzwiller approach
such that:%
\begin{equation}
\left\langle \Psi _{G}\left\vert \widehat{O}\right\vert \Psi
_{G}\right\rangle =\left\langle \Psi _{0}\left\vert \widehat{O}%
_{G}\right\vert \Psi _{0}\right\rangle ,  \label{GW}
\end{equation}%
where 
\begin{equation}
\left\langle \phi _{i\alpha }\left\vert \widehat{O}_{G}\right\vert \phi
_{j\beta }\right\rangle =z_{i\alpha }O_{i\alpha ,j\beta }z_{j\beta }
\end{equation}%
if $\left( i,\alpha \right) \neq \left( j,\beta \right) $ while 
\begin{equation}
\left\langle \phi _{i\alpha }\left\vert \widehat{O}_{G}\right\vert \phi
_{i\alpha }\right\rangle =O_{i\alpha ,i\alpha }.
\end{equation}%
Therefore we have 
\begin{eqnarray}
\left\langle \Psi _{G}\left\vert \widehat{O}\right\vert \Psi
_{G}\right\rangle &=&\sum_{i\alpha ,j\beta }^{\prime }z_{i\alpha }z_{j\beta
}O_{i\alpha ,j\beta }\left\langle \Psi _{0}\left\vert c_{i\alpha }^{\dagger
}c_{j\beta }\right\vert \Psi _{0}\right\rangle  \notag \\
&&+\sum_{i\alpha }O_{i\alpha ,i\alpha }\left\langle \Psi _{0}\left\vert
c_{i\alpha }^{\dagger }c_{i\alpha }\right\vert \Psi _{0}\right\rangle
\end{eqnarray}%
where $\left\{ \phi _{i\alpha }\right\} $ is a local orbital basis for the
system, a subset $L$ of which represents localized electrons in the system
and $\sum_{i\alpha ,j\beta }^{\prime }$ indicates summation with the self
term $\left( i,\alpha \right) =\left( j,\beta \right) $ omitted. $\left\vert
\Psi _{0}\right\rangle $ is the uncorrelated Hartree-like wavefunction
corresponding to $\left\vert \Psi _{G}\right\rangle $. The $z$-factors are
renormalization weights for the localized part of the one-particle density
matrix%
\begin{equation}
z_{i\alpha }=\frac{\sum_{\Gamma _{i},\Gamma _{i}^{\prime }}\sqrt{p_{\Gamma
_{i},\Gamma _{i}^{\prime }}^{\alpha }}}{\sqrt{n_{i\alpha }\left(
1-n_{i\alpha }\right) }}
\end{equation}%
where $p_{\Gamma ,\Gamma ^{\prime }}^{\alpha }=p_{i,\Gamma }p_{i,\Gamma
^{\prime }}\left\vert \left\langle \Gamma ^{\prime }\left\vert c_{i\alpha
}\right\vert \Gamma \right\rangle \right\vert ^{2}$ is the probability for a
transition between two atomic configurations $\Gamma _{i}$ and $\Gamma
_{i}^{\prime }$ that results in the increase of the occupation of the single
particle state $\alpha $ at site $i$ by one. The summation is over all
configurations on site $i$\cite{Dorin,Buenemann}. For non-localized orbitals 
$z_{i\alpha }=1$. In our present notation, $\alpha $ includes both the
orbital and spin indices and \ 
\begin{equation}
n_{i\alpha }=\left\langle \Psi _{0}\left\vert c_{i\alpha }^{\dagger
}c_{i\alpha }\right\vert \Psi _{0}\right\rangle .
\end{equation}%
The $z_{i\alpha }$ are therefore functions of the orbital occupation $%
\left\{ n_{i\alpha }\right\} $ and the probabilities $p_{\Gamma ,\Gamma
^{\prime }}^{\alpha }$. The $p_{\Gamma ,\Gamma ^{\prime }}^{\alpha }$ can be
expressed in terms of the probabilities $p_{i}\left( \Gamma \right) $ for a
local configuration $\Gamma $\cite{Dorin,Buenemann}. The set of local
orbitals in $L$ and the set of local configurations $\left\{ \Gamma
_{i}\right\} $ with non-zero probabilities are specified for the system.

Under the Gutzwiller approximation, the electron density is defined as 
\begin{eqnarray}
\rho \left( \mathbf{r}\right) &=&\sum_{i\alpha ,j\beta }^{\prime }z_{i\alpha
}z_{j\beta }\phi _{i\alpha }^{\ast }\left( \mathbf{r}\right) \phi _{j\beta
}\left( \mathbf{r}\right) \left\langle \Psi _{0}\left\vert c_{i\alpha
}^{\dagger }c_{j\beta }\right\vert \Psi _{0}\right\rangle  \notag \\
&&+\sum_{i\alpha }\left\vert \phi _{i\alpha }\left( \mathbf{r}\right)
\right\vert ^{2}\left\langle \Psi _{0}\left\vert c_{i\alpha }^{\dagger
}c_{i\alpha }\right\vert \Psi _{0}\right\rangle
\end{eqnarray}%
We can also define the localized electron density $\rho _{l}\left( \mathbf{r}%
\right) $ by a similar expression, except that the summation is restricted
to $\alpha $ and $\beta $ in $L$.

We will choose $E_{xc}\left[ \rho \right] $ to be of the form:%
\begin{eqnarray}
E_{xc}\left[ \rho \right] &=&\sum_{i\Gamma }p_{i}\left( \Gamma \right)
U_{\Gamma }+\int d^{3}r\left( \rho -\rho _{l}\right) \varepsilon _{xc}\left(
\rho \right)  \notag \\
&&-\frac{1}{2}\int \int \rho _{l}\left( \mathbf{r}\right) v\left( \mathbf{r,r%
}^{\prime }\right) \rho _{l}\left( \mathbf{r}^{\prime }\right)
d^{3}rd^{3}r^{\prime }  \label{xcgw}
\end{eqnarray}
We require our system to be the same as the regular LDA system in the limit
when there are no localized electrons. This can be achieved if we choose $%
\varepsilon _{xc}\left( \rho \right) $ to be the same as in LDA. In the
limit when all electrons are localized our system becomes a multiband
Hubbard Hamiltonian. $U_{\Gamma }$ is a sum of Slater integrals representing
the Coulomb repulsion between localized orbitals on the same site in the
configuration $\Gamma $.

The variational degrees of freedom in our system are $\left\{ p_{i}\left(
\Gamma \right) \right\} $ and $\left\vert \Psi _{0}\right\rangle $. Since $%
\left\vert \Psi _{0}\right\rangle $ can be expressed as a simple product of
one particle wavefunctions $\left\{ \psi _{n\mathbf{k}}\right\} $, it
follows that 
\begin{equation}
\left\langle \Psi _{0}\left\vert c_{i\alpha }^{\dagger }c_{j\beta
}\right\vert \Psi _{0}\right\rangle =\sum_{n,\mathbf{k}}f_{n,\mathbf{k}%
}\left\langle \psi _{n\mathbf{k}}|\phi _{i\alpha }\right\rangle \left\langle
\phi _{j\beta }|\psi _{n\mathbf{k}}\right\rangle
\end{equation}%
where $m,\mathbf{k}$ are the usual band indices and $f_{n,\mathbf{k}}$ is $1$
for occupied states and $0$ for empty states. The variational parameters in
our calculations are $\left\{ p_{i}\left( \Gamma \right) \right\} $ and $%
\left\{ \psi _{n\mathbf{k}}\right\} $ with the constraints that $\left\{
\psi _{n\mathbf{k}}\right\} $ are normalized to $1$.

A set of single electron equations can be derived using the variational
principle by taking the derivatives of the new energy functional Eq.\ref%
{funct} with respect to $\left\{ \psi _{n\mathbf{k}}\right\} $ and $\left\{
p_{i}\left( \Gamma \right) \right\} $, keeping in mind that the density $%
\rho \left( \mathbf{r}\right) $, the localized density $\rho _{l}\left( 
\mathbf{r}\right) $, the exchange-correlation functional $\varepsilon
_{xc}\left( \rho \right) $and the parameters $z_{i\alpha }$ are defined
above. This set of equations can be solved self-consistently to give the
band structures and total energies of the correlated electron system.

By taking the derivatives with respect to $\left\{ \psi _{n\mathbf{k}%
}\right\} $ we have%
\begin{equation}
\widehat{H}_{\mathrm{eff}}\psi _{n\mathbf{k}}=\lambda _{n\mathbf{k}}\psi _{n%
\mathbf{k}}  \label{Schr}
\end{equation}%
with effective Hamiltonian%
\begin{equation}
\widehat{H}_{\mathrm{eff}}=\widehat{H}_{G}^{l}+\sum_{i\alpha \beta
}2e_{i\beta }\frac{\partial \ln z_{i\beta }}{\partial n_{i\alpha }}\widehat{P%
}_{i\alpha }  \label{Heff}
\end{equation}%
where the first term $\widehat{H}_{G}^{l}$ is the Gutzwiller-renormalized
operator of $\widehat{H}^{l}$ where 
\begin{equation}
\widehat{H}^{l}=\widehat{T}+\widehat{V}_{\mathrm{ion}}+\widehat{V}_{H}+\mu _{%
\mathrm{xc}}-\widehat{P}_{l}\left( \widehat{V}_{H}^{l}+\varepsilon _{\mathrm{%
xc}}\right) \widehat{P}_{l}
\end{equation}%
is the effective mean-field potential with the localized-localized
electron-interaction contributions subtracted out. $\widehat{V}_{H}$ and $%
\widehat{V}_{H}^{l}$ are the mean field Coulomb potential (Hartree
potential) due to the total and localized charge respectively. 
\begin{equation}
\mu _{\mathrm{xc}}=\frac{\partial \left( \rho -\rho _{l}\right) \varepsilon
_{xc}\left( \rho \right) }{\partial \rho }
\end{equation}%
%
%
% is the exchange-correlation potential
while $\widehat{P}_{l}$ is the projection operator on the localized subspace
\ $L$, i.e. $\widehat{P}_{l}\phi _{i\alpha }=\phi _{i\alpha }$ for $\alpha
\in L$ (localized electron orbitals) and $\widehat{P}_{l}\phi _{i\alpha }=0$
otherwise.

The second term in Eq.\ref{Heff} adds back the localized-localized electron
contribution to the effective potential (subtracted from the first term)
according to the Gutzwiller approximation. In the second term, $\widehat{P}%
_{i\alpha }$ is the projection operator on $\phi _{i\alpha }$ and 
\begin{equation}
e_{i\alpha }=\frac{1}{2}\sum_{n\mathbf{k}}f_{n\mathbf{k}}\left\langle \psi
_{n\mathbf{k}}\left\vert \widehat{P}_{i\alpha }\widehat{H}_{G}^{l}+\widehat{H%
}_{G}^{l}\widehat{P}_{i\alpha }\right\vert \psi _{n\mathbf{k}}\right\rangle -%
\widehat{H}_{i\alpha ,i\alpha }^{l}n_{i\alpha }
\end{equation}%
The derivatives over the local configuration probabilities $\left\{
p_{i}\left( \Gamma \right) \right\} $ yield%
\begin{equation}
0=U_{\Gamma }+2\sum_{i\alpha }e_{i\alpha }\frac{\partial \ln z_{i\alpha }}{%
\partial p_{i}\left( \Gamma \right) }
\end{equation}%
This is a set of self-consistency criteria to be satisfied by$\left\{
p_{i}\left( \Gamma \right) \right\} $.

The set of equations in (13) and (18) can be solved iteratively to obtain a
self-consistent solution for $\left\{ p_{i}\left( \Gamma \right) \right\} $
and $\left\{ \psi _{n\mathbf{k}}\right\} $ and the total energy of the
system evaluated according to equations (1) to (12).

The specific derivation of our density functional approach is based on the
Gutzwiller method. It is, however, possible to formally generalize the
approach and set up a density functional for an arbitrary solvable
interacting many-electron reference system, independent of the specific
aspects of the Gutzwiller formalism. To put it in a more general context, we
discuss these aspects of our approach. A crucial ingredient of the density
functional formulation by Kohn and Sham is to choose the kinetic energy
functional $T\left[ \rho \right] $ to be the kinetic energy of a system of
independent electrons in a potential $\widehat{V}_{\mathrm{s}}$that yields
the ground state density $\rho \left( \mathbf{r}\right) $. Our approach
differs from this key starting point by Kohn and Sham by defining $T\left[
\rho \right] $ as the kinetic energy of a system of\textit{\ interacting}
electrons. We then assume that one can again formulate an effective many
body problem with Hamiltonian 
\begin{equation}
\widehat{H}_{\mathrm{s}}=\widehat{T}+\widehat{V}_{\mathrm{s}}+\widehat{U}_{%
\mathrm{s}}
\end{equation}%
that yields the ground state density $\rho \left( \mathbf{r}\right) $
through an appropriate choice of the single particle potential $\widehat{V}_{%
\text{s}}$. $\widehat{U}_{\text{s}}$ still contains explicit interactions
among the electrons. We then use the wave function $\left\vert \Psi _{\text{s%
}}\right\rangle $ of this correlated reference system to define the kinetic
energy functional 
\begin{equation}
T\left[ \rho \right] =\left\langle \Psi _{\mathrm{s}}\left\vert \widehat{T}%
\right\vert \Psi _{\mathrm{s}}\right\rangle .
\end{equation}%
This necessitates a new, modified functional for $E_{xc}\left[ \rho \right] $
which is no longer the only term where correlation effects enter. A
self-consistent set of density functional equations emerges for any choice
of $\left\vert \Psi _{\text{s}}\right\rangle $ that allows for an evaluation
of the functional derivative $\delta T\left[ \rho \right] /\delta \rho
\left( \mathbf{r}\right) $. The Gutzwiller wave function $\left\vert \Psi _{%
\text{s}}\right\rangle =$ $\left\vert \Psi _{G}\right\rangle $ discussed
above or other Jastrow type wave functions are examples. The close formal
connection to a non-interacting electron system (see Eq.\ref{GW}) makes the
analysis of the kinetic energy feasible and leads to the mapping of the
many-particle problem onto a set of effective single particle Schr\"{o}%
dinger equations.

We can make further progress in our analysis of $E_{xc}\left[ \rho \right] $
by using the coupling constant integration approach of Ref.\cite{Harris74}.
We first make a specific choice $\widehat{U}_{\mathrm{s}}=\widehat{P}%
\widehat{U}\widehat{P}$ for the interaction term, where $\widehat{P}$
projects onto the configuration space of the strongly interacting electrons
e.g. those among local $3d$, $4f$ or $5f$ electrons. We then introduce the
Hamiltonian 
\begin{equation}
\widehat{H}_{\lambda }=\widehat{T}+\widehat{U}_{\mathrm{s}}+\lambda \left( 
\widehat{U}-\widehat{U}_{\mathrm{s}}\right) +\widehat{V}\left( \lambda
\right) .
\end{equation}%
with varying coupling constant $\lambda $, where the potential $\widehat{V}%
\left( \lambda \right) $ equals the nuclear potential $\widehat{V}_{0}$ for $%
\lambda =1$ and is assumed to yield a density $\rho \left( \mathbf{r}\right) 
$ independent of $\lambda $ for $\lambda <1$. With the help of the
Hellmann-Feynman theorem we obtain an explicit expression for the exchange
correlation functional: 
\begin{eqnarray}
E_{xc}\left[ \rho \right] &=&\left\langle \Psi _{\mathrm{s}}\left\vert 
\widehat{U}_{\mathrm{s}}\right\vert \Psi _{\mathrm{s}}\right\rangle +\int
d^{3}r\left( \rho \varepsilon _{xc}\left( \mathbf{r}\right) -\rho
_{l}\varepsilon _{xc,l}\left( \mathbf{r}\right) \right)  \notag \\
&&-\frac{1}{2}\int \int \rho _{l}\left( \mathbf{r}\right) v\left( \mathbf{r,r%
}^{\prime }\right) \rho _{l}\left( \mathbf{r}^{\prime }\right)
d^{3}rd^{3}r^{\prime },  \label{xcgen2}
\end{eqnarray}%
that is independent of the specifics of the above Gutzwiller approach. The
exchange correlation potential $\varepsilon _{xc,l}\left( \rho \right) $ of
the localized orbitals is found to be 
\begin{equation}
\varepsilon _{xc,l}\left( \mathbf{r}\right) =\frac{1}{2}\int d^{3}r^{\prime
}\rho _{l}\left( \mathbf{r}^{\prime }\right) v\left( \mathbf{r,r}^{\prime
}\right) \left( g_{l}\left( \mathbf{r,r}^{\prime }\right) -1\right) ,
\end{equation}%
where $g_{l}\left( \mathbf{r,r}^{\prime }\right) =\int_{0}^{1}d\lambda \frac{%
\sum_{\sigma \sigma ^{\prime }}\left\langle \widehat{P}b_{\sigma \sigma
^{\prime }}^{\dagger }\left( \mathbf{r,r}\right) b_{\sigma \sigma ^{\prime
}}\left( \mathbf{r,r}\right) \widehat{P}\right\rangle _{\lambda }}{\rho
_{l}\left( \mathbf{r}\right) \rho _{l}\left( \mathbf{r}^{\prime }\right) }$ $%
\ $\ is the two particle correlation function of localized states with $%
b_{\sigma \sigma ^{\prime }}\left( \mathbf{r,r}\right) =\psi _{\sigma
^{\prime }}\left( \mathbf{r}^{\prime }\right) \psi _{\sigma }\left( \mathbf{r%
}\right) $. Comparing this result with Eq.\ref{xcgw} we identify $%
\sum_{i\Gamma }p_{i}\left( \Gamma \right) U_{\Gamma }=\left\langle \Psi _{%
\mathrm{s}}\left\vert \widehat{U}_{\mathrm{s}}\right\vert \Psi _{\mathrm{s}%
}\right\rangle $ and find that we made the approximate choice $\varepsilon
_{xc,l}\left( \rho \right) \simeq \varepsilon _{xc}\left( \rho \right) $.
Then, the second term in Eq.\ref{xcgen2} becomes simply $\int d^{3}r\left(
\rho -\rho _{l}\right) \varepsilon _{xc}\left( \rho \right) $. This simple
choice reproduces correctly the limit of the ordinary density functional
theory of Kohn and Sham in the case without localized orbitals. A similar
form was also shown to be very successful in describing the non-linear
exchange-correlation interactions between core and valence charge densities%
\cite{louie82}.

An important physical consequence of our new approach is its ability to
combine two, seemingly distinct, mechanisms for screening the Coulomb
interaction between electrons. Usually, screening is understood as a
response of the particle density in the vicinity of a charged object and
many aspects of it are appropriately incorporated in the usual Kohn-Sham
density functional formalism. On the other hand, in case of strong
electron-electron correlations, the Coulomb interaction can also be screened
via a reorganization of the many body state, for example in the form of
strong band renormalizations, amounting to a drastic change in the kinetic
energy of the electrons. These effects are at best poorly described in the
usual density functional formalism. In our approach, through the
self-consistent solution for $\left\{ \psi _{n\mathbf{k}}\right\} $ and $%
\left\{ p_{i}\left( \Gamma \right) \right\} $, the system can respond to the
addition of extra terms in the Hamiltonian (both in the external potential
as well as \ in $U_{\Gamma }$). The approach has the advantage of combining
both aspects of screening in a self consistent way without resorting to
model parameters or model Hamiltonians. This opens a new perspective for the
first principles description of strong electron correlations in complex
materials.

In summary, we have developed a new density functional theory incorporating
strongly correlated electronic effects into the kinetic energy functional
within the Gutzwiller approximation. We show that a set of single particle
equations can be obtained from functional derivatives of the energy with
respect to the orbitals included in the non-interacting one-particle density
matrix with renormalized kinetic and potential operators. This set of
equations can be solved self-consistently in a way similar to regular LDA
calculations. In our scheme, the variational parameters are the single
particle electron wavefunctions for the localized and delocalized electrons
and the occupancy of the various localized configurations at each atom in
the unit cell. Like LDA, and unlike the many other correlated electron
calculations on the market, the formulation is from first principles with
all Coulomb and exchange-correlation interactions determined
self-consistently. There are no adjustable parameters. We believe
developments along this approach will be fruitful in extending the
successful applications of density functional calculations to new systems
with important electron correlations.

We are grateful to V. Antropov, B. N. Harmon, W. Weber and J. Buenemann for
useful discussions. Ames Laboratory is operated for the U.S. Department of
Energy by Iowa State University under Contract No. DE-AC02-07CH11358. This
work was supported by the Director for Energy Research, Office of Basic
Energy Sciences.

\end{document}